\documentclass[fleqn]{comjnl}
\usepackage{amsfonts}
\usepackage{amsmath}
\usepackage{amssymb}
\usepackage{semantic}
\usepackage{xspace}
\usepackage{algorithm}
\usepackage{algpseudocode}
\usepackage{xcolor}
\usepackage{listings}
\usepackage{url}
\usepackage{comment}
\usepackage{tabularx}
\usepackage{soul}
\usepackage{hyperref}

\newtheorem{note}{Note}

\catcode`\|=\active
\def|{\mid}

\renewcommand{\comp}{\circ}
\newcommand{\inv}[1]{#1^\smallsmile}
\newcommand{\cmpt}[1]{#1^{-}}
\newcommand{\fun}{\rightarrow}
\DeclareMathSymbol{\dres}{\mathbin}{AMSa}{"43}
\DeclareMathSymbol{\rres}{\mathbin}{AMSa}{"42}
\def \ndres     {\mathbin{\rlap{\raise.05ex\hbox{$-$}}{\dres}}}
\def \nrres     {\mathbin{\rlap{\raise.05ex\hbox{$-$}}{\rres}}}
\def \dom       {\mathop{\mathrm{dom}}}
\def \id        {\mathop{\mathrm{id}}}
\def \ran       {\mathop{\mathrm{ran}}}
\def \defs      {\mathrel{\widehat=}}
\newcommand{\e}{\emptyset}
\newcommand{\plus}{\mathbin{\scriptstyle\sqcup}}
\newcommand{\w}{\{\cdot \plus \cdot\}}
\newcommand{\p}{(\cdot,\cdot)}
\newcommand{\Disj}{\parallel}
\newcommand{\true}{\mathit{true}}
\newcommand{\false}{\mathit{false}}

\newcommand{\lfun}{\rightarrow}

\renewcommand{\Cup}{\mathit{un}}
\renewcommand{\Cap}{\mathit{inters}}

\newcommand{\disj}{\Disj}
\newcommand{\Comp}{\mathit{comp}}

\newcommand{\Inv}{\mathit{inv}}
\newcommand{\Id}{\mathit{id}}
\newcommand{\set}{\mathit{set}}
\newcommand{\Nset}{\mathit{nset}}
\newcommand{\pair}{\mathit{pair}}
\newcommand{\Npair}{\mathit{npair}}
\newcommand{\Rel}{\mathit{rel}}
\newcommand{\Nrel}{\mathit{nrel}}

\newcommand{\Dom}{\mathit{dom}}
\newcommand{\Ran}{\mathit{ran}}
\newcommand{\Ncup}{\mathit{nun}}

\newcommand{\Nin}{\notin}

\newcommand{\Apply}{\mathit{apply}}
\newcommand{\Pfun}{\mathit{pfun}}
\newcommand{\Ncap}{\mathit{ninters}}

\newcommand{\setlog}{$\{log\}$\xspace}
\newcommand{\LBR}{\mathcal{L}_\mathcal{BR}}
\newcommand{\SATREL}{\mathit{SAT}_\mathcal{BR}\xspace}

\newcommand{\FSTRA}{\textsf{FS\&RA}\xspace}
\newcommand{\Var}{\mathcal{V}}
\newcommand{\Ur}{\mathcal{X}}
\newcommand{\FUr}{\mathcal{F}_\Ur}
\newcommand{\PiSB}{\Pi_\mathit{S}}
\newcommand{\PiRB}{\Pi_\mathit{R}}
\newcommand{\sU}{\mathsf{O}}
\newcommand{\sSet}{\mathsf{Set}}

\newcommand{\iS}{\mathcal{R}}
\newcommand{\iF}[1]{(#1)^\iS}
\newcommand{\sS}{\mathsf{S}}

\newcommand{\Theorem}{\mathsf{theorem}}
\newcommand{\Assume}{\mathsf{assume}}
\newcommand{\Prove}{\mathsf{prove}}
\newcommand{\Drop}{\mathsf{drop}}
\newcommand{\Cases}{\mathsf{cases}}
\newcommand{\Rewrite}{\mathsf{rewrite}}

\newcommand{\Define}{\mathsf{define}}

\newcommand{\Counterex}{\mathsf{counterex}}

\definecolor{dkblue}{rgb}{0,0.1,0.5}
\definecolor{lightblue}{rgb}{0,0.5,0.5}
\definecolor{dkgreen}{rgb}{0,0.4,0}
\definecolor{dk2green}{rgb}{0.4,0,0}
\definecolor{dkpink}{rgb}{1,0,1}
\lstdefinelanguage{SSR}{
mathescape=true,
texcl=false,
morekeywords=[1]{From, Section, Module, End, Require, Import, Export, Defensive, Function, Variable, Variables, Parameter, Parameters, Axiom, Hypothesis, Hypotheses, Notation, Local, Tactic, Reserved, Scope, Open, Close, Bind, Delimit, Definition, Let, Ltac, Fixpoint, CoFixpoint, Add, Morphism, Relation, Implicit, Arguments, Set, Unset, Contextual, Strict, Prenex, Implicits, Inductive, CoInductive, Record, Structure, Canonical, Coercion, Theorem, Lemma, Corollary, Proposition, Fact, Remark, Example, Proof, Goal, Save, Qed, Defined, Hint, Resolve, Rewrite, View, Search, Show, Print, Printing, All, Graph, Projections, inside, outside, Locate, Maximal},
morekeywords=[2]{forall, exists, exists2, fun, fix, cofix, struct, match, with, end, as, in, return, let, if, is, then, else, for, of, nosimpl},
morekeywords=[3]{Type, Prop},
morekeywords=[4]{pose, set, move, case, elim, apply, clear, hnf, intro, intros, generalize, rename, pattern, after, destruct, induction, using, refine, inversion, injection, rewrite, congr, unlock, compute, ring, field, replace, fold, unfold, change, cutrewrite, simpl, have, gen, generally, suff, wlog, suffices, without, loss, nat_norm, assert, cut, trivial, revert, bool_congr, nat_congr, abstract, symmetry, transitivity, auto, split, left, right, autorewrite},
morekeywords=[5]{by, done, exact, reflexivity, tauto, romega, omega, assumption, solve, contradiction, discriminate},
morekeywords=[6]{do, last, first, try, idtac, repeat},
literate={isn't }{{{\ttfamily\color{dkgreen} isn't }}}1,
showstringspaces=false,
morestring=[b]",
morestring=[d]',
tabsize=2,
extendedchars=true,
sensitive=true,
breaklines=true,
basicstyle=\ttfamily,
captionpos=b,
columns=[l]fullflexible,
identifierstyle={\ttfamily\color{black}},
keywordstyle=[1]{\ttfamily\color{blue}},
keywordstyle=[2]{\ttfamily\color{dkgreen}},
keywordstyle=[3]{\ttfamily\color{lightblue}},
keywordstyle=[4]{\ttfamily\color{dkblue}},
keywordstyle=[5]{\ttfamily\color{red}},
keywordstyle=[6]{\ttfamily\color{dkpink}},
stringstyle=\ttfamily,
commentstyle=\rmfamily,
}
\let\C=\lstinline

\begin{document}

\title{Proof Automation in the Theory of Finite Sets and Finite Set Relation Algebra}

\author{Maximiliano Cristi\'a}
\affiliation{Universidad Nacional de Rosario and CIFASIS, Argentina}\email{cristia@cifasis-conicet.gov.ar}

\author{Ricardo D. Katz}
\affiliation{CIFASIS-CONICET, Argentina} \email{katz@cifasis-conicet.gov.ar}

\author{Gianfranco Rossi}
\affiliation{Universit\`a di Parma, Italy}\email{cristia@cifasis-conicet.gov.ar, katz@cifasis-conicet.gov.ar, gianfranco.rossi@unipr.it}

\shortauthors{Cristi\'a, Katz and Rossi}


\begin{abstract}
\setlog (`setlog') is a satisfiability solver for formulas of the theory of
finite sets and finite set relation algebra (\FSTRA). As such, it can be used
as an automated theorem prover (ATP) for this theory. \setlog is able to
automatically prove a number of \FSTRA theorems, but not all of them.
Nevertheless, we have observed that many theorems that \setlog cannot
automatically prove can be divided into a few subgoals automatically
dischargeable by \setlog. The purpose of this work is to present a prototype
interactive theorem prover (ITP), called \setlog-ITP, providing evidence that a
proper integration of \setlog into world-class ITP's can deliver a great deal
of proof automation concerning \FSTRA. An empirical evaluation based on 210
theorems from the TPTP and Coq's SSReflect libraries shows a noticeable
reduction in the size and complexity of the proofs with respect to Coq.
\end{abstract}

\maketitle

%%%%%%%%%%%%%%%%%%%%%%%%%%%%%%%%%%%%%%%%%%%%%%%%%%%%%%%%%%%%%%%%%%%%
\section{Introduction}
\label{sec:intro}
%%%%%%%%%%%%%%%%%%%%%%%%%%%%%%%%%%%%%%%%%%%%%%%%%%%%%%%%%%%%%%%%%%%%

Interactive theorem proving (ITP) \cite{DBLP:series/hhl/HarrisonUW14} is
increasingly used in the formalization and proof of results of mathematics and
logic, and also a widely used approach to formal verification of hardware and
software. ITP's such as Coq \cite{DBLP:series/txtcs/BertotC04}, Isabelle/HOL
\cite{DBLP:books/sp/NipkowPW02} and HOL Light \cite{DBLP:conf/fmcad/Harrison96}
vary in the level of expressivity and automation, but typically support rich
specification languages including higher-order logic or dependent type theory.
Since interactive theorem proving is labor intensive, thus costly and limited,
much research has been devoted to the development of automated reasoning.

SMT solvers \cite{Nieuwenhuis00} implement decision procedures for the
satisfiability problem of formulas of specific theories. Since they have
significantly improved their power in the last decades, it has become
increasingly common for ITP's the use of SMT solvers as efficient automated
theorem provers (ATP) for the corresponding theories. As a consequence, users
of mainstream ITP's can call ATP's
\cite{DBLP:journals/jar/CzajkaK18,DBLP:conf/lpar/PaulsonB10} and SMT solvers
\cite{DBLP:conf/cav/EkiciMTKKRB17,DBLP:journals/jar/BlanchetteBP13} to
automatically advance their proofs. These ad-ons exploit the idea of mixing
interactive and automated proof steps. Our proposal fits in this line of work
as we propose to integrate \setlog as a special purpose ATP into ITP's.

%The formal methods community has made breakthrough advances concerning the automation of formal proofs. SMT solvers can be used as efficient ATP's for a variety of theories. However, only recently, some SMT solvers are able to natively deal with a fragment of set theory and binary relations \cite{DBLP:conf/cade/MengRTB17}. Traditional ATP's such as E prover \cite{DBLP:journals/aicom/Schulz02} and Vampire \cite{DBLP:journals/aicom/RiazanovV02} can efficiently solve many FOL problems. Users of mainstream ITP's can call ATP's \cite{DBLP:journals/jar/CzajkaK18,DBLP:conf/lpar/PaulsonB10} and SMT solvers \cite{DBLP:conf/cav/EkiciMTKKRB17,DBLP:journals/jar/BlanchetteBP13} to automatically advance their proofs. These ad-ons exploit the idea of mixing interactive and automated proof steps. Our proposal fits in this line of work as we propose to integrate \setlog as a special purpose ATP into ITP's.

\setlog is a satisfiability solver accepting an input language at least as
expressive as the class of full set relation algebras on finite sets (\FSTRA)
\cite{DBLP:journals/jar/CristiaR20}. \FSTRA essentially corresponds to the
first-order fragment of formal notations such as Alloy \cite{Jackson00}, B
\cite{Abrial00} and Z \cite{Spivey00} restricted to finite sets. In
consequence, this input language can be used as a specification language for a
large class of software systems and \setlog as a tool to reason about them.

\setlog can automatically prove\footnote{By an abuse of language, from now on
we will say that \setlog can or cannot prove a theorem to mean that it can
decide or not the satisfiability of its negation (see Remark~\ref{RemarkNegation} below).} 97\% of the theorems on Boolean algebra (BOO),
relation algebra (REL) and set theory (SET) gathered in the TPTP library
\cite{Sut09} that can be expressed in its input language (in .3 s each in
average), see \cite{DBLP:conf/RelMiCS/CristiaR18}. Since the equational theory
of \FSTRA is undecidable \cite{andreka1997decision}, \setlog cannot decide the
satisfiability of all the formulas it accepts. Moreover, \setlog can take too
long to decide the satisfiability of some formulas. For example, it takes a
long time to prove the following result:
\begin{equation}\label{eq:571}
\begin{split}
f \in{} & A \fun B  {}\land{} \ran f = B {}\land{} g \in B \fun C \\ {}\land{}
& h \in B \fun C \land
f \comp g = f \comp h
\implies g = h
\end{split}\tag{T1}
\end{equation}
where $A$, $B$ and $C$ denote any finite sets, $A \fun B$ denotes the set of
all (finite) functions from $A$ to $B$ (in this context a function is a binary
relation where no two ordered pairs have the same first component) and $\ran f$
is the range of $f$. Nevertheless, since $g = h \iff g \subseteq h \land h
\subseteq g$, the proof of \eqref{eq:571} can be reduced to the proofs of the
following two implications, on which \setlog spends a few seconds:
\begin{gather*}
\begin{split}
  f \in{} & \star \fun \star {}\land{} \ran f = B {}\land{} g \in B \fun \star \\ {}\land{}
&  h \in \star \fun \star \land
  f \comp g = f \comp h
  \implies g \subseteq h
\end{split} \label{eq:571_1} \\
\begin{split}
  f \in{} & \star \fun \star {}\land{} \ran f = B {}\land{} g \in \star \fun \star \\ {}\land{}
&  h \in B \fun \star \land
  f \comp g = f \comp h
  \implies h \subseteq g
\end{split} \label{eq:571_2}
\end{gather*}
Here $\star$ means that the corresponding hypotheses can be dropped (for
example, $f \in \star \fun \star$ says that it does not matter what the domain
and range of $f$ are). Thus, by \emph{dividing} the proof of \eqref{eq:571}
into subgoals and by \emph{dropping} the appropriate hypotheses in each,
\setlog can automatically do the rest.

We have noticed that in practice the approach above succeeds on proving many
results that \setlog cannot automatically prove. As a consequence, in this
paper we present \setlog-ITP, a prototype interactive theorem prover where
users can enter any \setlog formula and interactively prove it. More precisely,
in \setlog-ITP users can divide the proof into subgoals, drop hypotheses and
call \setlog to perform the actual mathematical steps. This follows the
way other tools, such as Atelier B \cite{Mentre00} and the Coq's why3
\cite{Bobot00} tactic, work. We point out that \setlog-ITP is just a vehicle to
provide evidence that a proper integration of \setlog's rewriting system into
world-class ITP's can deliver a great deal of proof automation concerning
\FSTRA.

%RK As opposed to the usual fact that the more hypothesis are available during
%an interactive proof, the better, dropping hypothesis is decisive to the success
%of the proposed approach. Indeed, when an ATP is called to perform a proof step,
%unnecessary hypothesis may make it walk through many actually useless proof
%paths (and, in the case of incomplete tools like \setlog, they can take an
%infinite  proof path). Hence, by dropping unnecessary hypothesis the prover has
%fewer proof paths to walk through, thus augmenting the chances to end the proof
%and to do it faster.

In order to validate our proposal in practice we 
perform a number of proofs with \setlog-ITP and Coq.
%RK the latter with and without the use of CoqHammer \cite{DBLP:journals/jar/CzajkaK18}.
On these theorems \setlog either does not terminate or takes a very long time
to do it. This comparison shows promising results as Coq proofs are harder and
longer than \setlog-ITP's
%RK, while those using CoqHammer are less automatic
(see Section~\ref{emp} for details).

As SMT solvers, \setlog generates a solution when it determines that a formula
is satisfiable. Indeed, \setlog provides a finite representation of all the
(possibly infinitely many) solutions \cite{DBLP:journals/jar/CristiaR20}. In
the context of integrating \setlog into an ITP, this means that if \setlog is
called to advance a proof but it happens that the goal is not provable from the
premises, a counterexample is generated. This counterexample might help the
user to adjust the theorem or the theory which contains it. QuickChick has been
proposed as a tool to decrease the number of failed proof attempts in Coq by
generating counterexamples before a proof is attempted
\cite{denes2014quickchick}. This tool relies on a random counterexample
generator. Although \setlog counterexamples are generated in a deductive
fashion, it is also more limited as it works only for a specific theory.

Let us finally mention that traditional ATP's such as E prover
\cite{DBLP:journals/aicom/Schulz02} and Vampire
\cite{DBLP:journals/aicom/RiazanovV02} can automate proofs of \FSTRA. Since
they can efficiently solve many FOL problems, they work by encoding set theory
in (most often) untyped first-order logic. One of the simplest encodings
applies extensionality and rewrites away all definitions, thus arriving at
formulas based on set membership. However, these encodings must deal with
typing information when sets do not have the same set support. The easiest way
to deal with this issue is to omit all type information, but this approach is
unsound. Another way to deal with types is to annotate terms with type tags or
guards. This considerably increases the size of the problems passed to generic
ATP, with a dramatic impact on their performance
\cite{DBLP:journals/corr/BlanchetteB0S16,DBLP:conf/lpar/BuryDDHH15}. Another
approach to proof automation in set theory is to use polymorphic provers. Our
work follows this approach as \setlog can be seen as a specialized prover for
polymorphic set theory. The empirical assessment presented in this paper
confirms the results reported by other polymorphic provers
\cite{DBLP:conf/lpar/BuryDDHH15,DBLP:conf/asm/BuryCDE18,altergo,ArchSat,Zipperposition}.

This paper is structured as follows. The logic language supported by \setlog
and some of its main features are presented in Section~\ref{sec:setlog}.
Section~\ref{itp} describes \setlog-ITP which is empirically evaluated in
Section~\ref{emp}. We
%RKdiscuss our approach in Section \ref{relwork} and
give our final conclusions in Section~\ref{concl}.

%%%%%%%%%%%%%%%%%%%%%%%%%%%%%%%%%%%%%%%%%%%%%%%%%%%%%%%%%%%%%%%%%%%%
\section{\setlog: a satisfiability solver for finite sets and  relations}
\label{sec:setlog}
%%%%%%%%%%%%%%%%%%%%%%%%%%%%%%%%%%%%%%%%%%%%%%%%%%%%%%%%%%%%%%%%%%%%

In this section we provide a brief, informal introduction to the \setlog system \cite{setlog}. A formal presentation of \setlog's language can be found in \ref{app:setlog}; deeper
presentations can be found elsewhere \cite{DBLP:journals/jar/CristiaR20,DBLP:conf/RelMiCS/CristiaR18,DBLP:conf/cav/CristiaR16}.

\subsection{The \setlog language}
\label{syntax}

\setlog \cite{setlog} is a satisfiability solver implemented in Prolog whose
input language is denoted $\LBR$. This is a multi-sorted first-order predicate
language with two distinct sorts: the sort $\sSet$ of all the terms which
denote sets (including binary relations) and the sort $\sU$ of all the other
terms.
%and the sort $\Bool$ representing the two-valued domain of truth values
%$\{\mathsf{false},\mathsf{true}\}$. %Terms of either sort are allowed
%to enter in the formation of set terms (in this sense, the designated
%sets are hybrid), no nesting restrictions being enforced (in particular,
%membership chains of any finite length can be modeled).
%In a term which is not a variable and designates an ur-element, the main
%functor (be it a constant or a function symbol) will act as a free (`uninterpreted')
%Herbrand constructor; a special set constructor, and a handful of reserved
%predicate symbols endowed with a pre-designated set-theoretic meaning,
%are also available. Formulas are built in the usual way by using conjunction,
%disjunction and negation of atomic predicates. A number of complex operators
%(in the form of predicates) are defined as $\LBR$ formulas, thus making
%it simpler for the user to write complex formulas.
Thus, we do not introduce distinct sorts for sets and binary relations. Binary
relations are just sets of ordered pairs. This allows sets and relations to be
freely mixed; in particular all set operators are directly applicable to binary
relations.

\begin{note}[Background on binary relations]\label{note:background}
Let $R$ and $S$ be binary relations, and $A$ a set.
Then, we define following relational operators: relational composition,
      $R \comp S = \{(x,z) | \exists y ((x,y) \in R \land (y,z) \in S)\}$;
converse (or inverse) of $R$, $\inv{R} = \{(y,x) | (x,y) \in R\}$;
the identity relation on $A$, $\id A = \{(x,x) | x \in A\}$;
domain of $R$, $\dom R = \{x | \exists y ((x,y) \in R)\}$;
range of $R$, $\ran R = \dom \inv{R}$;
domain restriction of $R$ on $A$, $A \dres R = \id(A) \comp R$;
range restriction of $R$ on $A$, $R \rres A = R \comp \id(A)$;
domain anti-restriction of $R$ on $A$,
        $A \ndres R = R \setminus (A \dres R)$;
range anti-restriction of $R$ on $A$,
        $R \nrres A = R \setminus (R \rres A)$;
relational image of $A$ through $R$,
      $R[A] = \ran(A \dres R)$;
and relational overriding of $R$ by $S$,
      $R \oplus S = (\dom S \ndres R) \cup S$.
A binary relation is a (partial) function if no two of its ordered pairs
have the same first component. Given that functions are just binary relations,
all relational operators can be applied to functions.
%, and all set operators can be applied to both of them.}
      \qed
\end{note}

In $\LBR$ set operators are encoded as constraints over the domain of
finite hybrid sets. For example: $\Cup(A,B,C)$ is a constraint interpreted as
$C = A \cup B$. \setlog implements a wide range of set and relational operators
(cf. Note~\ref{note:background}). For instance, $\in$ is a constraint
interpreted as set membership; $=$ is set equality; $\Dom(R,D)$ corresponds to
$\dom R = D$; $A \subseteq B$ corresponds to the subset relation;
$\Comp(R,S,T)$ is interpreted as $T = R \comp S$; and $\Apply(F,X,Y)$ is
equivalent to $\Pfun(F) \land [X,Y] \in F$, where $\Pfun(F)$ constrains $F$ to
be a (partial) function. Formulas in \setlog are conjunctions ($\land$) and
disjunctions ($\lor$) of constraints. Negation is introduced
by means of so-called \emph{negated constraints}. For example $\Ncup(A,B,C)$ is
interpreted as $C \neq A \cup B$ and $\Nin$ as the negation of $\notin$. In
general, if $\pi$ is a constraint, $n\pi$ corresponds to its negated form. For
formulas to fit inside the decision procedures implemented in \setlog, users
must only use this form of negation.

In turn, terms can be either uninterpreted Herbrand terms (as in Prolog) or set
terms, i.e., terms with the following form and interpretation: $\emptyset$
to denote the empty set; $\{x \plus A\}$, called
\emph{extensional set}, which is interpreted as $\{x\} \cup A$, where $A$
is a set term; and $A \times B$ to
represent the Cartesian product between the sets denoted by set terms
$A$ and $B$.

In \setlog sets are always finite and untyped and they are allowed as set
elements (i.e., sets can be nested). As the second argument of an extensional
set can be a variable, sets in \setlog can be unbounded.

\begin{note}[Binary relations and expressivenes]\label{note:background}

$\LBR$ turns out to be at least as expressive as \emph{the class of full set
relation algebras on finite sets}. A \emph{full set relation algebra}
\cite{andreka1997decision} over a base set $A$, denoted $\mathfrak{R}(A)$, is
the relation algebra where relations are subsets of $A \times A$. A mapping
from formulas of a $\mathfrak{R}(A)$ with $A$ finite to $\LBR$ formulas can be
easily defined \cite{DBLP:journals/jar/CristiaR20}. The class of full set
relation algebras on finite sets is the class of relation algebras
$\mathfrak{R}(A)$ where $A$ is a finite set. Further, $\LBR$ allows for the
definition of the class of full set heterogeneous relation algebras on finite
sets. An heterogeneous relation algebra deals with relations between two
arbitrary sets $A$ and $B$, i.e. with subsets of $A \times B$
\cite{Schmidt1997}. We use the acronym \FSTRA to denote such a class, for any
$A$ and $B$ finite.

The class of full set heterogeneous relation algebras roughly corresponds to
the first-order fragment of formal notations such as Alloy \cite{Jackson00}, B
\cite{Abrial00} and Z \cite{Spivey00}. A large class of programs can be
specified within this fragment. The limitation of $\LBR$ to finite sets is not
so severe as many programs operate only on finite data structures. Therefore,
$\LBR$ can be used as a specification language for a large class of software
systems and \setlog as a tool to reason about them
\cite{CristiaRossiSEFM13,Cristi__2020,DBLP:journals/corr/abs-2009-00999,hcvs2017,CristiaRossiSETS14}.
\qed
\end{note}

%%%%%%%%%%%%%%%%%%%%%%%%%%%%%%%%%%%%%
\subsection{\label{rewriting}A rewriting system for $\LBR$}
%%%%%%%%%%%%%%%%%%%%%%%%%%%%%%%%%%%%%

\setlog implements a rewriting system for $\LBR$
formulas, called $\SATREL$, whose global organization is shown in Algorithm~\ref{glob}. Basically, $\SATREL$ uses two procedures: \textsf{sort\_infer} and
\textsf{STEP}.

\begin{algorithm}%[htbp]
\begin{algorithmic}[0]
 \State $\Phi \gets \textsf{sort\_infer}(\Phi)$;
% \Repeat
%   \State $\Phi'' \gets \Phi$;
   \Repeat
     \State $\Phi' \gets \Phi$;
     \State $\Phi \gets \textsf{STEP}(\Phi)$
   \Until{$\Phi = \Phi'$;}
%   \State $\Phi \gets \textsf{remove\_neq}(\Phi)$
%  \Until{$\Phi = \Phi''$;}
 \State\Return{$\Phi$}
\end{algorithmic}
\caption{The solver $\SATREL$. $\Phi$ is the input formula.} \label{glob}
\end{algorithm}

$\LBR$ does not provide variable declarations. For this reason,
\textsf{sort\_infer}($\Phi$) automatically adds either a $\set$ or a $\Rel$
constraint for each variable $X$ in $\Phi$ which is required to represent,
respectively, either a set or a binary relation according to the intended
interpretation of the terms or constraints where $X$ occurs.

\textsf{STEP} applies specialized rewriting procedures to the current formula
$\Phi$ and returns either $\false$ or a modified formula. Each rewriting
procedure applies a few non-deterministic rewrite rules which reduce the
syntactic complexity of $\LBR$
constraints of one kind. The execution of $\textsf{STEP}$ is iterated until a
fixpoint is reached, i.e., the formula is irreducible. \textsf{STEP} returns
$\false$ whenever (at least) one of the procedures in it rewrites $\Phi$ to
$\false$.

The rewriting procedures implemented in \setlog can be divided into two
classes: those for set constraints and those for relational constraints. The
former were introduced in \cite{Dovier00} and extensively discussed from then
on. They constitute the base for a decision procedure for finite sets based on
set unification and set constraint solving. The latter were introduced more
recently \cite{DBLP:journals/jar/CristiaR20,DBLP:conf/RelMiCS/CristiaR18}.
Roughly, there are 50 rewriting procedures adding up 175 rewrite rules.

\begin{figure}
\begin{gather}
\begin{split}
\{x  {}\plus{}& A\} = \{y \plus B\} \lfun \\
  & x = y \land A = B \\
  & \lor x = y \land \{x \plus A\} = B \\
  & \lor x = y \land A = \{y \plus B\} \\
  & \lor A = \{y \plus N\} \land \{x \plus N\} = B
\end{split} \label{rule:eq} \\
\begin{split}
\Cup & (A,B,\{t \plus C\}) \lfun \\
     & \{t \plus C\} = \{t \plus N\} \land t \notin N \\
     & \land (A = \{t \plus N_1\} \land \Cup(N_1,B,N) \\ %\land t \notin N_1
     & {}\qquad\lor B = \{t \plus N_1\} \land \Cup(A,N_1,N) \\
     & {}\qquad\begin{split}
        {}\lor A &= \{t \plus N_1\} \\ % & \land t \notin N_1 \\
                                 & \land B = \{t \plus N_2\}
                               %   \land t \notin N_2
                                   \land \Cup(N_1,N_2,N))
        \end{split}   \\
\end{split} \label{rule:cup} \\
\begin{split}
\Inv&(R,\{(y,x) \plus S\}) \rightarrow \\
& R = \{(x,y) \plus N\} \land \Inv(N,S)
 \end{split}  \label{rule:inv}
%
%\begin{split}
%    \Comp&(\{(x,t) \plus R\},\{(u,z) \plus S\},T) \rightarrow \\
%           & \Comp(\{(x,t)\},\{(u,z)\},N_1)\\
%           & \land \Comp(\{(x,t)\},S,N_2) \\
%           & \land \Comp(R,\{(u,z)\},N_3)\\
%           & \land \Comp(R, S, N_4) \\
%           & \land \Cup(N_1,N_2,N_3,N_4,T)
%\end{split} \label{rule:comp}
\end{gather}
\caption{Representative rewriting rules}
\label{fig:rr}
\end{figure}

Here we just show some of the most representative rewrite rules in
Figure~\ref{fig:rr} (the reader can find a comprehensive list online
\cite{calculusBR}). Note that these rules are recursive. Rule \eqref{rule:eq}
finds all possible solutions for the equality between two non-empty extensional
sets. The second and third disjuncts take care of duplicates in the right- and
left-hand side terms, respectively, while the last disjunct takes care of
permutativity of the set constructor $\w$. Specifically, the last disjunct can
be read as `$y$ must belong to $A$, $x$ must belong to $B$ and there exists a
set $N$ containing the remaining elements of both $A$ and $B$'. In turn,
rule \eqref{rule:cup} finds all possible solutions of a set union operation
when the result is a non-empty extensional set, where $N$, $N_1$ and $N_2$ are
new variables (implicitly existentially quantified). Note that set unification
is used to avoid possible repetitions of $t$ in $C$. Also observe that the
disjunction captures the three possible solutions: $t$ belongs to $A$, $t$
belongs to $B$ and $t$ belongs to $A$ and $B$. Finally, rule \eqref{rule:inv}
finds the binary relation whose converse is a non-empty extensional relation in
a very simple way.
%RK Finally, rule~\eqref{rule:comp} applies when $T$ is a variable in
%which case it computes the composition between two non-empty relations.
%This rule is based on one of the axioms of relation algebra which states
%that $(A \cup B) \comp C = (A \comp C) \cup (B \comp C$).

The rewriting system implemented by \setlog has been proved to be a
semi-decision procedure \cite{DBLP:journals/jar/CristiaR20}. More precisely, it
has been proved that: a) when Algorithm~\ref{glob} terminates, the returned
formula preserves the set of solutions of the input formula; b) the returned
formula is $\false$ if and only if the input formula is unsatisfiable; and c)
if the returned formula is not $\false$ it is trivial to calculate one of its
solutions (basically by substituting all set variables by the empty set). In
this context, a `solution' is an assignment of values to all the free variables
of the formula. Furthermore, when \setlog terminates, it has the ability to
produce a finite representation of all the (possibly infinitely many) solutions
of the input formula, in the form of a finite disjunction of $\LBR$ formulas.
In other words, whenever \setlog terminates, it either produces a proof of
unsatisfiability or a finite representation of all the solutions.

%%%%%%%%%%%%%%%%%%%%%%%%%%%%%%%%%%%%%
\subsection{\label{using}Using \setlog}
%%%%%%%%%%%%%%%%%%%%%%%%%%%%%%%%%%%%%

Users interact with \setlog by simply entering a formula; there are no user
commands. If the formula is unsatisfiable \setlog will simply return $\false$
and if it is satisfiable it will return a finite representation of all its
solutions.

\begin{example}
For example, the following is a satisfiable formula (note that binary
relations can be freely combined with extensional sets, and set operators can
take relations as arguments):
\begin{equation}\label{e:ex1}
\begin{split}
\Cup&(A,B,\{(1,1), (h,3) \plus C \times D\}) \\
& \land \Id(E,A) \land \Inv(B,B) \land 1 \notin E
\end{split}
\end{equation}
% un(A,B,{[1,1], [H,3] / C}) & id(X,A) & inv(B,B) & 1 nin X
% un(A,B,{[1,1], [H,3] / cp(C,D)}) & id(E,A) & inv(B,B) & 1 nin E
The relevant part of a solution returned by \setlog is:
\begin{gather*}
A = \{(3,3) \plus N_3\},
B = \{(1,1) \plus N_2\}, \\
h = 3,
E = \{3 \plus N_1\} \\
\textsf{Constraint: } 3 \notin C, un(N_2,N_3,C \times D), id(N_1,N_3), \\
\qquad \qquad \; \; \; inv(N_2,N_2), \dots
\end{gather*}
where $N_i$ are fresh variables. That is, each solution is composed of a
(possibly empty) conjunction of equalities between variables and terms and a
(possibly empty) conjunction of constraints. The conjunction of constraints is
guaranteed to be trivially satisfiable. \qed
\end{example}

In this context, \setlog can be used as a set-based, constraint-based
programming language. Users can give values to what they consider to be input
variables in the formula and \setlog will return values for the remaining
variables. For instance, if \eqref{e:ex1} is thought as a program where $A$ and
$B$ are inputs and the user enters \eqref{e:ex1} conjoined with $A =
\{(2,2),(3,3)\} \land B = \{(1,1),(1,2),(2,1)\}$, the answer will be $h = 3, C
= D = \{1,2\}, E = \{2,3\}$. In \setlog, \emph{formulas are programs}.
% A = {[2,2],[3,3]} & B = {[1,1],[1,2],[2,1]} &
% un(A,B,{[1,1], [H,3] / cp(C,D)}) & id(E,A) & inv(B,B) & 1 nin E

Given that \setlog is a satisfiability solver we can use it also as an
automated theorem prover. To prove that formula $\Phi$ is a theorem, \setlog
has to be called on $\lnot \Phi$ waiting an $\false$ answer, meaning that
$\lnot \Phi$ is unsatisfiable (and thus $\Phi$ is a theorem).

\begin{example}
We can prove that set intersection is commutative by asking \setlog to prove
the following formula is unsatisfiable:
\[
\Cap(A,B,C) \land \Cap(B,A,D) \land C \neq D
\]
As there are no finite sets satisfying this formula, \setlog returns $\false$.
The formula can also be written as:
\[
\Cap(A,B,C) \land \Ncap(B,A,C) \hfill\qed
\]
\end{example}

All these properties along with programming facilities not discussed in this
paper \cite{setlog}, make \setlog a versatile verification tool
\cite{CristiaRossiSEFM13,Cristi__2020,DBLP:journals/corr/abs-2009-00999,hcvs2017,CristiaRossiSETS14}.

%%%%%%%%%%%%%%%%%%%%%%%%%%%%%%%%%%%%%
%\section{\label{itp}Implementing an ITP on top of \setlog}
\section{Automating complex \FSTRA proofs}\label{itp}
%%%%%%%%%%%%%%%%%%%%%%%%%%%%%%%%%%%%%
As we have pointed out, \setlog may not terminate or may take a very long time when it is used to prove some theorems of \FSTRA. However, we have observed that in practice the proofs of many of such theorems can be divided into a few subgoals each of which can be automatically and quickly discharged by \setlog. In fact, these proofs follow a recurring pattern: divide the proof by introducing some assumptions, drop zero or more hypotheses in each subgoal and call \setlog to do the hard, annoying mathematical work. This would imply that complex theorems of \FSTRA can be easily proved by calling \setlog at the right points.

In order to provide evidence of these observations, we have developed a proof-of-concept ITP on top of \setlog that we call \setlog-ITP. This is a freely available +500 LOC Prolog program \cite{setlogITP} that allows users to declare a \setlog formula as a theorem and attempt to prove it interactively through some proof commands. First, we will show a typical proof using \setlog-ITP, and then we will give technical details about the proof system.

\begin{remark}[Limitations]
It is important to bear in mind that \setlog-ITP is intended to be an ITP \emph{only} for theorems of \FSTRA and \emph{only} to empirically validate our proposal. This means that it cannot be compared in no way with general-purpose ITP's such as Coq or Isabelle/HOL.
% For instance, \setlog-ITP users will not be able to axiomatize some mathematical theory and then develop a library of results on it. They will only be able to prove results on \FSTRA. In other words, \setlog-ITP complements \setlog by allowing interactive proofs when automated \setlog proofs are infeasible or unpractical.
\end{remark}

\subsection{A typical proof}\label{exampleproof}

\begin{figure*}
\quad \quad \quad \quad \quad \quad \quad \quad \quad \quad \quad
{\footnotesize
\inference[{\footnotesize$\Rewrite$}]%
  {
     \inference[{\footnotesize$\Drop$}]
         {\inference[{\footnotesize$\Prove$}]{\setlog}{\Gamma_1 \vdash g \subseteq h}}
         {\Gamma \vdash g \subseteq h}
       &
         \inference[{\footnotesize$\Drop$}]
         {\inference[{\footnotesize$\Prove$}]{\setlog}{\Gamma_2 \vdash g \subseteq h}}
         {\Gamma \vdash h \subseteq g}
  }
  {\Gamma \vdash g = h}
}
\caption{\label{fig:t1proof} \setlog-ITP proof of theorem $\mathsf{T1}$ or \eqref{eq:571}}
\end{figure*}

\eqref{eq:571} is declared as a \setlog-ITP theorem with the $\Theorem$ command:
\begin{equation*}
\begin{split}
\Theorem(&\mathsf{T1}, \\
         & \Pfun(f) \land
           f \subseteq A \times B \land
           \Dom(f,A) \land
           \Ran(f,B) \\
         & \land \Pfun(g)
           \land g \subseteq B \times C
           \land \Dom(g,B) \\
         & \land \Pfun(h)
           \land h \subseteq B \times C   \\
         & \land \Dom(h,B) \land \Comp(f,g,N) \land \Comp(f,h,N), \\
         & g = h)
\end{split}
\end{equation*}
where the first parameter is just a name for the theorem, the second one is a (possibly empty) conjunction of hypotheses and the third one is the thesis, both entered as \setlog formulas. Note that $f \in A \fun B$ is encoded in \setlog as $\Pfun(f) \land f \subseteq A \times B \land \Dom(f,A)$ and that $f \comp g = f \comp h$ is encoded as two $\Comp$ constraints yielding the same result ($N$).

As we have said, an automated proof of \eqref{eq:571} would take a long time. However, the interactive proof shown in Figure~\ref{fig:t1proof} takes only a few seconds of computing time. There, $\Gamma$ represents the hypotheses of \eqref{eq:571}. The proof starts with the $\Rewrite$ command
which splits the proof into the two subgoals shown in Figure~\ref{fig:t1proof}. Attempting to use \setlog to prove these subgoals by means of the command $\Prove$ would consume as much time as the proof of the initial goal because they are essentially the same. As the proof of these two subgoals is symmetric, we will explain in detail only the first one. In this case the user can use the following $\Drop$ command, which expects a list of \setlog constraints:
% (in the implementation the command waits for a list of hypothesis identifiers):
\begin{equation}\label{eq:drop}
\begin{split}
\Drop([&f \subseteq A \times B, \Dom(f,A), \\
&      g \subseteq B \times C,
      h \subseteq B \times C, \Dom(h,B)])
\end{split}
\end{equation}
These constraints are expected to be part of $\Gamma$ in which case they are removed from it, thus yielding the following hypotheses:
\[
\begin{split}
&\Pfun(f) \land \Ran(f,B) \land \Pfun(g) \land \Dom(g,B) \\
& \land \Pfun(h) \land \Comp(f,g,N) \land \Comp(f,h,N)
\end{split}
\]
called $\Gamma_1$ in Figure~\ref{fig:t1proof}. Now, $\Prove$ discharges the current subgoal in a few seconds.
Since in this case \setlog succeeds in proving the goal, the system shows to the user the remaining subgoal.
A similar course of action is taken to discharge the remaining subgoal, where a different list of constraints is passed in to the $\Drop$ command. 

In Section~\ref{Sec_Discussion} we discuss some aspects of this proof and present the complete proof script in \eqref{script571}.

\subsection{Proof commands}

\begin{figure*}
\quad \quad \quad \quad \quad \quad
\begin{tabular}{ccc}
\inference[$\Assume(\varphi)$]%
  {\Gamma,\varphi \vdash \Delta & \Gamma \vdash \varphi}%
  {\Gamma \vdash \Delta}
&
\inference[$\Cases$]%
  {\Gamma \vdash \varphi,\Delta & \Gamma \vdash \xi,\Delta}%
  {\Gamma \vdash \varphi \land \xi, \Delta}
&
\inference[$\Drop(\varphi)$]%
  {\Gamma \vdash \Delta}%
  {\Gamma,\varphi \vdash \Delta}     \\[5mm]
\inference[$\Define(\pi)$]
  {\Gamma, \pi(\dots,n) \vdash \Delta}
  {\Gamma \vdash \Delta}
&
\multicolumn{2}{c}{%
\inference[$\Prove$]
  {\mathsf{setlog}(\Gamma \implies \Delta)}
  {\Gamma \vdash \Delta}
}   \\[5mm]
\multicolumn{3}{c}{%
\inference[$\Rewrite$]%
  {\inference%
     {\inference%
        {\inference%
           {\Gamma, (v = t)[n/x] \vdash \varphi[n/x]}%
           {\Gamma \vdash (v = t \implies \varphi)[n/x]}
        }
        {\Gamma \vdash \forall x:v = t \implies \varphi}
      &
      \inference%
        {\inference%
           {\Gamma, (w = u)[n/x] \vdash \xi[n/x]}%
           {\Gamma \vdash (w = u \implies \xi)[n/x]}
        }
        {\Gamma \vdash \forall x:w = u \implies \xi}
     }
     {\Gamma \vdash (\forall x:v = t \implies \varphi)
                     \land (\forall x:w = u \implies \xi)}%
  }
  {\Gamma \vdash \pi(v,w)}
} \\
\end{tabular}
\caption{\label{fig:proofcmd}Main \setlog-ITP proof commands as inference rules}
\end{figure*}

In this section we present in detail the main proof commands of \setlog-ITP (see Figure~\ref{fig:proofcmd}). Some of them are direct implementations of well-known inference rules while others implement a few such rules in a single proof step.

\begin{remark}[Interfacing with \setlog]\label{RemarkNegation}
As we already said, as \setlog is a satisfiability solver, it can be used as an ATP.
Indeed, if \setlog finds that formula $\varphi$ is unsatisfiable in \FSTRA, then $\lnot\varphi$ is a theorem (in \FSTRA). %RK In other words, if \setlog is used as an ATP, users must enter the negation of the formulas they want to prove.
%RK This means that
Thus, when \setlog is used as a back-end system for an ITP, formulas must be negated before sending them from the ITP to \setlog. However, with the intention to simplify the presentation, we are not going to mention this negation process in the remaining of the paper. This implies that, for example, when in the command called $\Prove$ (Figure~\ref{fig:proofcmd}) we say that $\Gamma \implies \Delta$ is sent to \setlog it actually means that its negation $\Gamma \land \lnot \Delta$ is sent to it.
\qed
\end{remark}

Concerning Figure~\ref{fig:proofcmd}, the $\Assume$ command
%RK is the implementation of
can be seen as the implementation of a special case of
the Cut rule; $\Cases$ corresponds to conjunction introduction; and $\Drop$ is
a specialized version of
the Weakening rule where the antecedent is weakened in order to deliver to \setlog exactly the necessary hypotheses that yield the consequent.

$\Define$ waits for a constraint $\pi \in \{\Cup, \Inv, \Id, \Comp\}$. The last
argument of $\pi$ is expected to be a new variable and all the others must be
variables in the current scope (of the proper sort). For instance, if in the
current scope $R$ is a binary relation then the user can issue
$\Define(\Inv(R,S))$, where $S$ is a new variable, in which case $\Inv(R,S)$ is
added as a new hypothesis. This is sound because what we are doing is no more
than asserting that the converse of $R$ is called $S$. Now the user can make
assumptions on $S$. For example, $\Assume(\Pfun(S))$, which means that $S$
(i.e., the converse of $R$) is a function. Without such a command it would be
impossible to  consider assumptions on expressions assembled from variables in
the current scope, as in \setlog-ITP set and relational operations are
represented as predicates.

The $\Prove$ command simply calls \setlog on the current subgoal. In this case, there are three possible behaviors: \emph{a}) \setlog answers that
%RK $\Gamma \implies \Delta$
the current subgoal is indeed valid and so it is proved and the next one (if any) is shown to the user; \emph{b}) \setlog answers that there is a counterexample
%RK for the formula
(for instance if too many hypotheses have been dropped) in which case a proper error message is printed; and \emph{c}) \setlog takes too long and the user decides to interrupt the command. In \emph{b} and \emph{c} the proof is unchanged. In \emph{b} users can execute command $\Counterex$ to get a counterexample witnessing why the goal failed (recall the discussion on the generation of counterexamples in the introduction).

$\Rewrite$ calls \setlog to rewrite the thesis; that is \setlog is called to apply a rewrite rule to the thesis (eg. one of the rules of Figure~\ref{fig:rr}) thus generating one or more new goals to prove---this last case occurs when the rewrite rule is nondeterministic. The thesis is assumed to be a single constraint. Each of these new goals should be simpler to prove than the orginal one. For each new goal generated by the rewrite rule, $\Rewrite$ performs three proof steps in one (see Figure~\ref{fig:proofcmd}):
\begin{enumerate}
\item\label{1} It applies conjunction introduction to the goal. If the subgoal is a conjunction of constraints, then the user will prove one after the other.
\item It applies universal introduction on each subgoal generated in step~\ref{1}. As the new goal will in general be a universally quantified formula, these quantified variables are `introduced'.
\item It applies conditional introduction on each subgoal generated in step~\ref{1}. As in the previous step, the new goal will in general be a conditional and so hypotheses are `introduced' as well.
\end{enumerate}
For example, if $\Gamma \vdash \Pfun(f)$ is the current goal, the net effect of $\Rewrite$ is shown in Figure~\ref{f:rw}, where $x$, $y$, $z$, $v$ and $N$ are new variables. If, for instance, \setlog does not terminate on $\Gamma \vdash \Pfun(f)$, after $\Rewrite$ the user has two simpler goals to prove. In particular, the one on the right most often can be automatically and quickly discharged with $\Prove$.
In Figure~\ref{f:rw} $\Rewrite$ produces those two subgoals because in $\LBR$ we have:
\[
\begin{split}
&\Pfun(f) \Leftrightarrow \\
&\qquad (\forall v: v \in f \Rightarrow \pair(v)) \\
&\qquad  \land (\forall x,y,z: (x,y) \in f \land (x,z) \in f \Rightarrow y = z)
\end{split}
\]
where $v \in f$ is equivalent to $f = \{v \plus N\}$ for some new variable $N$, which yields the equalities seen in Figure~\ref{f:rw}. Then, when conjunction, universal quantification and implication are introduced as in Figure~\ref{fig:proofcmd}, we get the result shown in Figure~\ref{f:rw}.

\begin{figure*}
\begin{equation*}\label{eq:rwpfun}
\quad \quad \quad \quad \quad \quad \quad
\inference%[{\tiny$\Rewrite$}]%
  {\Gamma, f = \{(x,y),(x,z) \plus N\} \vdash y = z
   &
   \Gamma, f = \{v \plus N\} \vdash \pair(v)}
  {\Gamma \vdash \Pfun(f)}
\end{equation*}
\caption{\label{f:rw}Example of a rewrite step}
\end{figure*}

Actually, the inference rule given for $\Rewrite$ in Figure~\ref{fig:proofcmd}
is a simplification of the real rule. Here we assume that the current thesis is
a constraint $\pi$ depending on two variables ($v$ and $w$), which when
rewritten by \setlog delivers a conjunction of two universal formulas (in the
real case it can be any number of them). These formulas have all the same form
$ \mathit{equalities} \implies \mathit{predicate} $, where
$\mathit{equalities}$ is a (possibly empty) conjunction of equalities of the
form $var = term$ where $var$ is one of the variables on which $\pi$ depends
on; and $\mathit{predicate}$ is a (possibly empty) conjunction of $\LBR$
constraints. In Figure~\ref{fig:proofcmd} we assume that the
$\mathit{equalities}$ in each conjunct have exactly one equality.

It is important to remark that \setlog-ITP does not need a command, for
instance, to perform equality substitution because this is performed by \setlog
when $\Prove$ is executed. In effect, if $\Prove$ is issued, for example, on
$\Gamma, f = \{v \plus N\} \vdash \pair(v)$, then \setlog will substitute $f$
by $\{v \plus N\}$ in $\Gamma$.

%%%%%%%%%%%%%%%%%%%%%%%%%%%%%%%%%%%%%
\subsection{Discussion}\label{Sec_Discussion}
%%%%%%%%%%%%%%%%%%%%%%%%%%%%%%%%%%%%%

As can be seen in Figure~\ref{fig:proofcmd}, many \setlog-ITP's proof commands correspond to standard inference rules present in ITP's. Then, they can be easily replaced by the proof commands present in a particular ITP.

As concerns proof automation, the $\Drop$ command plays a
central role. In effect, it allows to call \setlog with the minimal set of
hypotheses as to prove the goal. This implies a reduction of the proof term and
consequently of the computing time.
As opposed to the usual fact that the more hypotheses are available during an interactive proof, the better, dropping hypotheses is decisive to the success of our approach. Indeed, when an ATP is called to perform a proof step, unnecessary hypotheses may make it walk through many actually useless proof paths (and, in the case of  tools like \setlog, that might not terminate in some cases, they can take an infinite  proof path). Hence, by dropping unnecessary hypotheses the prover has fewer proof paths to walk through, thus augmenting the chances to end the proof and to do it faster.

On the downside, the key role of $\Drop$ sensibly changes the proof style as now the user must determine which hypotheses are superfluous to prove a subgoal instead of using them to prove it.

We also note that our approach tends to reduce the influence of a good \emph{lemma engineering}. That is, users usually plan which lemmas go first and which follow, so as to use the former in the proofs of the latter. For example, the Coq proof of \eqref{eq:571} is the following\footnote{The reader does not need to understand the proofs, just to have an idea of their length and complexity.}:
\begin{lstlisting}[language=SSR]
move=>is_function_F is_function_G is_function_H range_F_eq_B rel_comp_eq.
apply/eqP; rewrite eqEsubset; apply/andP; split.
- exact: (auxT is_function_F is_function_G range_F_eq_B rel_comp_eq).
- symmetry in rel_comp_eq.
  exact: (auxT is_function_F is_function_H range_F_eq_B rel_comp_eq).
\end{lstlisting}
\noindent where \verb+auxT+ is a helper lemma whose importance was made evident after the first proof attempt \eqref{eq:571}. Indeed, \verb+auxT+ states that $g \subseteq h$ holds if the hypotheses of \eqref{eq:571} are satisfied. Its proof is the following:
\begin{lstlisting}[language=SSR]
move=>[fun_cond_F _ _] [_ domain_G_eq_B _] range_F_eq_domain_G rel_comp_eq;
apply/subsetP => p p_in_G.
have: p.1 \in range F.
  rewrite range_F_eq_domain_G -domain_G_eq_B.
  apply/in_domainP; exists p.2.
  by rewrite -[(p.1,p.2)]surjective_pairing.
move=> /in_range_restP [a [_ pair_in_F]].
have: (a,p.2) \in rel_comp F H.
  by rewrite -rel_comp_eq;apply/in_rel_compP; exists p.1; split;
    [exact: pair_in_F |
     rewrite -[(p.1,p.2)]surjective_pairing].
move=> /in_rel_compP [b [in_F in_H]].
have p1_eq_b: p.1 = b by apply:
((((fun_cond_F (a,p.1)) (a,b)) pair_in_F) in_F).
by rewrite [p]surjective_pairing p1_eq_b.
\end{lstlisting}

Now, compare Coq's proof of \eqref{eq:571} with \setlog-ITP's (cf. Section~\ref{exampleproof}):
\begin{gather}
\Rewrite. \label{script571} \\
\begin{split}
-\; & \Drop([f \subseteq A \times B, \Dom(f,A), \\
      &\qquad g \subseteq B \times C, \Dom(g,B),
      h \subseteq B \times C]), \\
& \Prove.
\end{split} \notag \\
\begin{split}
-\; & \Drop([f \subseteq A \times B, \Dom(f,A), \\
    &\qquad g \subseteq B \times C,
      h \subseteq B \times C, \Dom(h,B)]), \\
& \Prove.
\end{split} \notag
\end{gather}
where no helper lemma is necessary and the complex proof of \verb+auxT+ is
replaced by dropping hypotheses and then calling \setlog. On the other
hand, as in Coq, the user still needs to guide the proof by splitting the initial
goal into $g \subseteq h$ and $h \subseteq g$
%RK (i.e. $\Rewrite$)
and the symmetry used in the Coq proof (i.e., \verb+symmetry+) is still present
in the \setlog-ITP proof, when symmetric $\Drop$ commands are executed. Hence,
were \setlog available in Coq the proof of \eqref{eq:571} would not need the
helper lemma and it would still be compact and semi-automatic.

However, there is still room for further automation. CoqHammer
\cite{DBLP:journals/jar/CzajkaK18}
% translates Coq logic into untyped first-order logic and
uses external ATPs to automate Coq proofs.
% Clearly, CoqHammer can prove
% theorems of theories far beyond the reach of \setlog.
CoqHammer helps in automating the proof of \eqref{eq:571}:

\begin{lstlisting}[language=SSR]
move=> is_function_F is_function_G is_function_H range_F_eq_B rel_comp_eq.
apply/eqP; rewrite eqEsubset; apply/andP; split.
- hammer.
- hammer.
\end{lstlisting}
\noindent where \verb+hammer+ needs lemma \verb+auxT+ to prove both subgoals. \emph{However,} \verb+hammer+ \emph{cannot prove} \verb+auxT+ \emph{automatically}. Then, were CoqHammer \emph{and} \setlog available, the Coq proof of \eqref{eq:571} could be \emph{almost} automatic: first use $\Drop$ and $\Prove$ to prove \verb+auxT+; and then use \verb+hammer+ to prove \eqref{eq:571} as above.

CoqHammer depends on a good lemma engineering. Conversely, \setlog does not depend on such engineering but it can only prove results of \FSTRA. A combination between a general tool like CoqHammer with specialized provers such as \setlog, seems to be a promising strategy towards proof automation.

%RK As SMT solvers, \setlog generates a counterexample (or solution) whenever a satisfiable formula is provided. Unlikely to SMT solvers, \setlog counterexamples represent a collection of ground solutions or `points'. Furthermore, \setlog generates a finite representation of all the (possibly infinitely many) ground solutions.
%RK As SMT solvers, \setlog generates a solution when it determines that a formula is satisfiable. Indeed, \setlog provides a finite representation of all the (possibly infinitely many) solutions \cite{Cristia2019}. In the context of integrating \setlog into an ITP, this means that if \setlog is called to advance a proof but it happens that the goal is not provable from the premises, a counterexample is generated (cf. command $\Counterex$). This counterexample might help the user to adjust the theorem or the theory which contains it. QuickChick has been proposed as a tool to decrease the number of failed proof attempts in Coq by generating counterexamples before a proof is attempted \cite{denes2014quickchick}. This tool, however, relies on a random counterexample generator. Although \setlog counterexamples are generated in a deductive fashion, it is also more limited as it works only for a specific theory.

%%%%%%%%%%%%%%%%%%%%%%%%%%%%%%%%%%%%%
\section{Empirical assessment}\label{emp}
%%%%%%%%%%%%%%%%%%%%%%%%%%%%%%%%%%%%%
As we have said, our intention is to provide evidence that integrating \setlog's
rewriting system into mainstream ITP's will yield a noticeable increment in
proof automation concerning \FSTRA. This is \setlog-ITP's single purpose. To
this end, we performed 210 proofs with \setlog-ITP and Coq in order to compare
their complexity and length. In an attempt to avoid as much as possible a bias
towards \setlog-ITP, 21 proofs correspond to problems listed in the REL and SET
collections of the TPTP library (\eqref{eq:571} is an example)
%RK . These problems are either undecidable or \setlog takes a very long time to solve them
which satisfy that \setlog either does not terminate or takes a very long time to do it when it is applied to prove them\footnote{\setlog automatically and quickly proves all the other problems of TPTP.SET and TPTP.REL expressible in its input language \cite{DBLP:journals/jar/CristiaR20}.}. The remaining 189 proofs correspond to lemmas included in Coq's SSReflect finite set library, \verb+finset+\footnote{The remaining 147 lemmas of \texttt{finset} are not
expressible in the input language of \setlog as they include operators such as generalized union or
powerset.} \cite{DBLP:journals/jfrea/GonthierM10}. We chose \verb+finset+ for
three reasons: \emph{a}) it has been designed as to make it easy to prove those
lemmas in Coq; \emph{b}) a fragment of \verb+finset+'s set theory keeps a clear
relationship with respect to \setlog's; and \emph{c}) it would provide evidence that
proof automation of real Coq results can be achieved with our proposal. Finally,
the Coq proofs were performed by one of the authors (with experience in working with SSReflect), while \setlog-ITP's were done by
another author.

As concerns the TPTP problems, they are encoded in Coq by extending SSReflect's
\verb+finset+. SSReflect is a proof language extending Coq with additional
tactics oriented to support long mathematical proofs. \verb+finset+ defines a
type for sets over a finite type. It includes definitions such as set
membership, union, Cartesian product, etc. However, it does not include set
relation algebra definitions such as the identity relation, converse (or
inverse), composition, and (partial) function. Since these are necessary to
reason about $\LBR$ formulas, we defined them in Coq. For example, our Coq
set-based definition of function from $A$ to $B$ (i.e., $f \in A \fun B$) is
the following:
\begin{lstlisting}[language=SSR]
Definition is_function_from_to
    (S1 S2:finType) (R:{set (S1 * S2)})
        (A:{set S1}) (B:{set S2}) :=
            [/\ is_function R, domain R = A
                & (range R) \subset B].
\end{lstlisting}
where \verb+is_function+, \verb+domain+ and \verb+range+ are defined in a similar fashion, but where \verb+\subset+ is part of SSReflect's finite type interface (on which \verb+finset+ is based on). We also give a set-based definition of the relational composition of two binary relations:
\catcode`\|=12
\begin{lstlisting}[language=SSR]
Definition rel_comp (S1 S2 S3:finType)
(R1:{set (S1 * S2)}) (R2:{set (S2 * S3)}) :=
    [set p | [ exists u, ((p.1, u) \in R1)
        && ((u, p.2) \in R2)]].
\end{lstlisting}
These extensions to \verb+finset+ lead to the following encoding of theorem \eqref{eq:571}:

\begin{lstlisting}[language=SSR]
Theorem T1 A B C (F:{set (S1 * S2)})
    (G H:{set (S2 * S3)}) :
        is_function_from_to F A B ->
        is_function_from_to G B C ->
        is_function_from_to H B C ->
        range F = B ->
        rel_comp F G = rel_comp F H -> G = H.
\end{lstlisting}

A similar encoding was used to state the 189 lemmas of \verb+finset+, as \setlog-ITP theorems. For example, the following \verb+finset+ lemma:
\begin{lstlisting}[language=SSR]
Lemma setCU A B :
    ~: (A :|: B) = ~: A :&: ~: B
\end{lstlisting}
where \verb+~:+ is complement ($\cmpt{}$), \verb+:|:+ is $\cup$ and \verb+:&:+ is $\cap$, is encoded as the following \setlog-ITP theorem:
\begin{equation*}
\begin{split}
\Theorem(& \text{\textsf{setCU}}, \\
  & A \subseteq T \land B \subseteq T \land \Cup(A,B,M_1) \\
  & \land \Cup(M_1,M_2,T) \land M_1 \disj M_2 \land \Cup(A,M_3,T) \\
  & \land A \disj M_3 \land \Cup(B,M_4,T) \land B \disj M_4, \\
  & \Cap(M_3,M_4,M_2))
\end{split}
\end{equation*}
where $T$ corresponds to the variable \verb+T+ of type \verb+finType+ declared
in the section containing \C$Lemma setCU$ (i.e., all sets of this section are
subsets of \verb+T+).

Table~\ref{t:experiments} summarizes the results of the evaluation. As we have said, \setlog is unable to automatically prove any of the 21 TPTP theorems (thus, for the TPTP theorems, the \textsc{Auto} entry is set to zero). On the other hand, Coq needs 690 proof commands to prove them, while
\setlog-ITP can do it with 219, of which only 146 are other than the $\Prove$
command. This is a reduction of about 68\% in the number of proof commands (and
about 79\% if $\Prove$ is not counted). By `proof command' we understand
all the characters accommodated in the same line
(which intend to represent basic mathematical proof steps). For example, for us, this is a
single Coq proof command:
\texttt{by move=> a; apply/setP=> x; rewrite inE; case: eqP => ->}.
In this sense, \setlog-ITP proof commands tend to be
simpler than Coq's. Actually, the 690 proof commands used in Coq to prove the
TPTP problems are composed of about 1030 applications of SSReflect tactics, which represent 35,014 characters while those used in \setlog-ITP just
3,662 characters. Then, apart from the gain in the number of proof commands,
there is notable gain in their complexity ($\approx$ 90\%).
Finally, collectively all the $\Prove$
commands consume 55 s of computing time which means that the automated part of
each theorem is executed in 2.6 s in average. On the other hand, the completion of all the Coq proofs took around 10 man-hour; while completing all the \setlog-ITP took around 1 man-hour.

\begin{table*}
\centering
\tabcolsep=5pt
\begin{tabularx}{\textwidth}{Xrrrrrrrr}
\hline
\multicolumn{1}{c}{\textsc{Collection}}
& \multicolumn{1}{c}{\textsc{\#}}
& \multicolumn{1}{c}{\textsc{Auto}}
& \multicolumn{1}{c}{\textsc{\%}}
& \multicolumn{3}{c}{\textsc{Commands}}
& \multicolumn{1}{c}{\textsc{Computing Time}}
& \multicolumn{1}{c}{\textsc{Avg Computing Time}} \\\cline{5-7}
 & & & & \multicolumn{1}{c}{\textsc{Coq}}
& \multicolumn{1}{c}{\textsc{\setlog}}
& \multicolumn{1}{c}{\textsc{non-$\Prove$}}
& & \\\hline
TPTP               &  21 &   0 &  0 & 690 & 219 & 146 &  55 s & 2.6 s \\
SSReflect \texttt{finset}
                   & 189 & 183 & 97 & 223 & 195 &   6 & 182 s & 1 s \\\hline
\textsc{Summary}   & 210 & 183 & 87 & 913 & 435 & 158 & 237 s & 1.1 s \\\hline
\end{tabularx}
\caption{\label{t:experiments}Summary of the empirical evaluation}
\end{table*}

As concerns the problems taken from \verb+finset+, Table~\ref{t:experiments}
indicates that \setlog-ITP automatically proves 97\% of them (in 1 s in
average). In this case the gain in the number of proof commands is minimal.
However, it should be noted that in \setlog-ITP 183 problems are proved via the
\emph{same} command ($\Prove$), while the Coq proofs require, roughly, 214
\emph{different} commands. In other words, a Coq user needs to reason on how to
prove 183 theorems, while a \setlog-ITP user does not. This fact can be
quantified if only the non-$\Prove$ commands are considered, as they amount to
only 2\% of the Coq commands.

The Coq proofs present in these experiments are the result of some degree of lemma engineering. For instance, in the TPTP Coq proofs we first proved 21 helper lemmas and then the 21 theorems used as experiment. The helper lemmas correspond to properties that are used several times in the proofs of the 21 theorems. These can be simple properties, such as the characterization of the fact that an ordered pair belongs to the relational composition of two binary relations, or more complex ones, such as the helper lemma \verb+auxT+ described in Section~\ref{Sec_Discussion}  (which is applied twice in the proof of \eqref{eq:571} thanks to the use of symmetry). Without this lemma engineering, proofs would be longer and more complex.
As usual, many of these helper lemmas make themselves evident after some of the main theorems are proved. In \setlog-ITP no lemma engineering was used (actually, there is no way to use or apply a previous lemma in the current proof). This is another indication of a gain in simplicity when \setlog-ITP is used.

This evaluation suggests that an integration of \setlog into Coq would produce more fully automated \FSTRA proofs and would semi-automate many others.

The full data set of our
experiments can be found online at:
\url{https://www.dropbox.com/s/c6z45thxlvr1q1h/setlogITP.zip?dl=0}.
They
%RK experiments
were performed on a Dell Latitude E7470 (06DC) laptop with a 4
core Intel(R) Core\texttrademark{} i7-6600U CPU at 2.60GHz with 8 Gb of main
memory, running Linux Ubuntu 18.04.2 LTS 64-bit with kernel 4.15.0-56-generic.
The following software versions were used: \setlog 4.9.6-5d over SWI-Prolog
(multi-threaded, 64 bits, version 7.6.4); Coq 8.9.0; CoqHammer 1.1 using Vampire
4.2.2, E prover 2.3, Z3 4.8.4.0 and CVC4 1.6.

%%%%%%%%%%%%%%%%%%%%%%%%%%%%%%%%%%%%%
\section{Conclusions}\label{concl}
%%%%%%%%%%%%%%%%%%%%%%%%%%%%%%%%%%%%%
We have proposed \setlog as a special purpose ATP for the theory of finite sets and finite set relation algebra, that can be integrated into ITPs to semi-automate proofs of this theory. We have also empirically evaluated the approach by implementing a prototype ITP where users can call \setlog as a proof command. The prototype was assessed on 210 theorems and compared
%RK to
with Coq. The assessment shows good results in: \emph{a}) the number of automated proofs; \emph{b}) the computing times needed to complete them; and \emph{c}) the reduction in the length and complexity of interactive proofs that call \setlog to discharge subgoals.

Concerning future work, in the case of Coq we see two possible integration strategies. The most obvious
one is to use \setlog as an external ATP, along the lines of SMTCoq \cite{DBLP:conf/cav/EkiciMTKKRB17} or CoqHammer \cite{DBLP:journals/jar/CzajkaK18}. In this case proof reconstruction might be
difficult or infeasible. %RK Then, we think in
A second integration strategy may consist in implementing \setlog's rewriting system as a Coq tactic. The first steps of this approach have already been done by Dubois and Weppe \cite{DBLP:conf/zum/DuboisW18}. Depending on the way this is done, as a side effect, this approach might yield a formal verification of
\setlog. It remains as open issues, though, whether the size of the proof term
produced by the new tactic will be more manageable than in the first strategy
and whether or not the tactic will be fast enough as to be worth it.

% decir que con un comando similar a apply algunas pruebas podrian reducirse aun mas

% decir que con un comando proveall o droprove se podria reducir mucho la longitud de las pruebas dado que muchos de los comandos son prove

%%
%% Bibliography
%%

%% Please use bibtex,

%biblio

%\bibliographystyle{compj}
%\bibliography{/home/mcristia/escritos/biblio}

%\end{document}

\appendix

%%%%%%%%%%%%%%%%%%%%%%%%%%%%%%%%%%%%%%%%%%%%%%%%%%%%%%%%%%%%%%%%%%%%
\section{Syntax and semantics of $\LBR$}
\label{app:setlog}
%%%%%%%%%%%%%%%%%%%%%%%%%%%%%%%%%%%%%%%%%%%%%%%%%%%%%%%%%%%%%%%%%%%%

In this appendix we provide a formal, detailed introduction of the syntax
and semantics of $\LBR$.

The input constraint language accepted by \setlog, $\LBR$, is a
first-order predicate language with terms of two sorts: terms designating sets
(including binary relations), and terms designating ur-elements. Terms of
either sort are allowed to enter in the formation of set terms (in this sense,
the designated sets are hybrid), no nesting restrictions being enforced (in
particular, membership chains of any finite length can be modeled). In a term
which is not a variable and designates an ur-element, the main functor (be it a
constant or a function symbol) will act as a free (`uninterpreted') Herbrand
constructor; a special set constructor, and a handful of reserved predicate
symbols endowed with a pre-designated set-theoretic meaning, are also
available. Formulas are built in the usual way by using conjunction,
disjunction and negation of atomic predicates. A number of complex operators
(in the form of predicates) are defined as $\LBR$ formulas, thus making it
simpler for the user to write complex formulas.

\subsection{Syntax}\label{app:syntax}

The syntax of $\LBR$ is defined primarily by giving the signature upon
which terms and formulas are built.

\begin{definition}%RK[Signature]
\label{signature}
The signature $\Sigma_{\cal BR}$ of $\LBR$ is a tuple $\langle
\mathcal{F},\Pi,\sSet,\sU,\Var\rangle$ where:
\begin{itemize}

\item $\mathcal{F}$ is the set of function symbols partitioned as
      $\mathcal{F} \defs \mathcal{F}_{S} \cup \FUr$, where
      $\mathcal{F}_{S} \defs \{\e,\w,\cdot \times \cdot\}$
      and $\FUr$ is a set of uninterpreted constant
      and function symbols, including at least the binary function
      symbol $\p$.

\item $\Pi$ is the set of predicate symbols partitioned as
$\Pi \defs \PiSB \cup \Pi_T \cup \PiRB$, where $\PiSB \defs
\{=,\neq,\in,\notin,\Cup,\disj\}$, $\Pi_T
\defs \{\set,\Nset,\Rel,\Nrel,\pair,\Npair\}$ and $\PiRB \defs \{\Id, \Comp,\Inv\}$.

\item $\{\sSet , \sU \}$ is the set of sorts.

\item $\Var$ is a denumerable set of variables partitioned as
$\Var \defs \Var_{S} \cup \Var_O$, where $\Var_{S}$ and
$\Var_O$ contain variables of sort $\sSet$ and $\sU$, respectively.
\qed
\end{itemize}
\end{definition}

%Intuitively, $\e$ represents the empty set; $\{ x \plus A\}$ represents the set
%$\{x\} \cup A$; $A \times B$ represents the Cartesian product of $A$ and $B$;
%$(x,y)$ represents an %ordered pair.
%RK  and $\Var_{S}$ andv$\Var_O$ represent sets of variables ranging over
%sets and ur-elements, respectively.

To complete the definition of $\LBR$, in addition to the signature it is
necessary to specify the {\em sorts} of function and predicate symbols: if $f
\in \mathcal{F}$ (resp., $\pi \in \Pi$) is of arity $n$, then its sort is an
$n+1$-tuple $\langle s_1, \ldots ,s_{n+1} \rangle$ (resp., an $n$-tuple
$\langle s_1, \ldots ,s_n \rangle$) of non-empty subsets of the set of sorts
$\{ \sSet , \sU \}$. This notion is denoted by $f:\langle s_1, \ldots
,s_{n+1}\rangle$ (resp., by $\pi:\langle s_1, \ldots ,s_n\rangle $).

\begin{definition}%RK[Sorts of function symbols]
\label{d:sorts}
The sorts of the function symbols in $\mathcal{F}$ are as follows:
\begin{itemize}
\item $\e: \langle \{\sSet \} \rangle$;
\item $\mathsf{\w: \langle \{\sSet , \sU\}, \{ \sSet \} , \{ \sSet\}\rangle }$;
\item $\cdot\times\cdot: \langle \{\sSet\}, \{ \sSet \}, \{ \sSet \}\rangle$;
\item $\p: \langle \{\sSet , \sU \} , \{ \sSet, \sU \} , \{ \sU \}\rangle$;
\item $f: \langle \{\sU\},\dots ,\{\sU\} \rangle \in (\{\sU\})^{n+1}$ if $f \in \FUr$ is of arity $n$.
\end{itemize}
The sorts of the predicate symbols in $\Pi$ are as follows
(symbols $=$, $\neq$, $\in$, $\notin$ and $\disj$ are infix; all other
symbols in $\Pi$ are prefix):
\begin{itemize}
\item $\pair, \Npair, \set, \Nset: \langle \{\sSet , \sU\} \rangle $;
\item $=,\neq: \langle \{\sSet , \sU) \}, \{ \sSet , \sU \} \rangle $;
\item $\in,\notin: \langle \{\sSet , \sU \} , \{\sSet \} \rangle $;
\item $\Cup, \Comp: \langle \{\sSet \} , \{\sSet \}, \{\sSet \} \rangle $;
\item $\disj, \Id, \Inv: \langle \{\sSet \} , \{\sSet \} \rangle $;
\item $\Rel, \Nrel: \langle \{\sSet \} \rangle $.
\qed
\end{itemize}
\end{definition}

We can now define the set of admissible (i.e., well-sorted) $\LBR$ terms.

\begin{definition}
All $\mathcal{BR}$-terms and their sorts are build inductively as follows:
\begin{itemize}
\item each variable $v \in V$ is a $\mathcal{BR}$-term of sort $\langle \{\sSet\} \rangle$ if $v \in \Var_{S}$ or sort $\langle \{\sU\}\rangle$ if $v \in \Var_{O}$.
\item if $f\in \mathcal{F}$ is a function symbol of sort $\langle s_1, \ldots , s_{n+1}\rangle$, and for each $i=1,\ldots , n$, $t_i$ is a $\mathcal{BR}$-term of sort $\langle s'_i\rangle$ with $s'_i \subseteq s_i$, then $f(t_1,\ldots ,t_n)$ is a $\mathcal{BR}$-term of sort $\langle s_{n+1}\rangle$.
\qed
\end{itemize}
\end{definition}

Note that the sort of any $\mathcal{BR}$-term $t$ is always of the form
$\langle \{\sSet\}\rangle$ or $\langle \{\sU\} \rangle$. In the former case we
simply say that $t$ is of sort $\sSet$, or a {\em set term}, and in the latter
case that $t$ is of sort $\sU$. In particular, $\mathcal{BR}$-terms of the form
$\w$ are called \emph{extensional} set terms. The first parameter of an
extensional set term is called \emph{element part} and the second is called
\emph{set part}. Observe that one can write terms representing sets which are
nested at any level.

The following notation is
introduced to make reading of set terms simpler: $\{t_1,t_2,\dots,t_n \plus
t\}$ as a shorthand for $\{t_1 \plus \{t_2 \,\plus\, \cdots \{ t_n \plus
t\}\cdots\}\}$ and the notation $\{t_1,t_2,\dots,t_n\}$ as a shorthand for
$\{t_1,t_2,\dots,t_n \plus \e\}$.

\begin{example}
The following are set terms:
\begin{itemize}
\item[-] $\e$
\item[-] $\{a,(b,c)\}$, i.e., $\{a \plus \{(b,c) \plus \e \}\}$, where $a$, $b$ and $c$ are constants of sort $\sU$
\item[-] $\{x \plus A \times \{y \plus B\}\}$, where $x,y$ are variables of sort $\sSet$ or $\sU$, and $A,B$ are variables of sort $\sSet$.
\item[-] $\{x \plus A\}$, where $x$ is a variable of sort $\sSet$ or $\sU$, and $A$ is a variable of sort $\sSet$.
\end{itemize}
On the opposite, $\{x \plus (a,b)\}$ is not a set term because $(a,b)$ is not of sort $\sSet$.
\qed
\end{example}

Finally, from $\LBR$ terms, we define $\LBR$ formulas.

\begin{definition}
All $\mathcal{BR}$-formulas are build inductively as follows:
\begin{itemize}
\item if $\pi\in \Pi$ is a predicate symbol of sort $\langle s_1, \ldots , s_n \rangle$, and for each $i=1,\ldots , n$, $t_i$ is a $\mathcal{BR}$-term of sort $\langle s'_i \rangle$ with $s'_i \subseteq s_i$, then $\pi (t_1,\ldots ,t_n)$ is a $\mathcal{BR}$-constraint, a particular case of $\mathcal{BR}$-formula.
\item if $\alpha$ and $\beta$ are $\mathcal{BR}$-formulas, then so are $\alpha \land \beta$ and $\alpha \lor \beta$.
\qed
\end{itemize}
\end{definition}

\begin{example}
The following are $\mathcal{BR}$-formulas:
\begin{gather*}
a \in A \land a \Nin B \land \Cup(A,B,C) \land C = \{x\} \\
\Cup(A,B,C) \land A \disj C \land \Inv(R,A) \land R \neq \e
\end{gather*}
where $a$ is a constant of sort $\sU$, $x$ is a variable of sort $\sSet$ or
$\sU$, and $A$, $B$, $C$ and $R$ are variables of sort $\sSet$. On the
contrary, $\Cup(A,B,(x,y))$ is not a $\mathcal{BR}$-formula because
$\Cup(A,B,(x,y))$ is not a $\mathcal{BR}$-constraint ($(x,y)$ is not of sort
$\sSet$ as required by the sort of $\Cup$). \qed
\end{example}

\subsection{\label{semantics}Semantics}

Semantics of $\mathcal{BR}$-formulas is given by defining
a suitable interpretation structure for $\LBR$.

Sorts and symbols in $\Sigma_{\cal BR}$ are interpreted according to the
interpretation structure $\iS \defs \langle D,\iF{\cdot}\rangle$, where $D$ and
$\iF{\cdot}$ are defined as follows.

\begin{definition}[Interpretation domain]
The interpretation domain $D$, of the interpretation structure $\iS$, is
partitioned as $D \defs D_\sSet \cup D_\sU$ where:
\begin{itemize}
\item $D_\sSet$ is the collection of all hereditarily finite hybrid
sets built from elements in $D$; and
\item $D_\sU$ is a collection of other objects, including ordered pairs of
elements in $D$.
\qed
\end{itemize}
\end{definition}

Hereditarily finite sets are those sets that admit (hereditarily finite) sets
as their elements. Note that, finite binary relations and functions, as defined
in Note~\ref{note:background}, belong to $D_\sSet$.

\begin{definition}[Interpretation function]
The interpretation function $\iF{\cdot}$, of the interpretation structure
$\iS$, is defined as follows.
\begin{itemize}
\item Each sort $\sS \in \{\mathsf{Set},\mathsf{O}\}$ is mapped to
      the domain $D_\sS$.
\item For each sort $ \sS\in \{\mathsf{Set},\mathsf{O}\}$, 
 each variable $x$ of sort $\sS$ is mapped to an element $x^\iS$ in  $D_\sS$.
\end{itemize}
The constant and function symbols in $\mathcal{F}_{S}$ are
interpreted as follows:
\begin{itemize}
\item $\e$ as the empty set;
\item $\{ x \plus A \}$ as the set $\{x^\iS\} \cup A^\iS$; and
\item $A \times B$ as the set $A^\iS \times B^\iS$.
\end{itemize}
The predicate symbols in $\Pi$ are interpreted as follows:
\begin{itemize}
\item $x = y$ as $x^\iS = y^\iS$;
\item $x \in A$ as $x^\iS \in A^\iS$;
\item $\Cup(A,B,C)$ as $C^\iS = A^\iS \cup B^\iS$;
\item $A \disj B$ as $A^\iS \cap B^\iS = \emptyset$;
\item $\set(x)$ as $x^\iS \in D_\sSet$;
\item $\pair(x)$ as $x^\iS \in \{(a,b) : a, b \in D\}$;
\item $\Rel(R)$ as $R^\iS \subset \{(a,b) : a, b \in D\}$;
\item $\Id(A,R)$ as $R^\iS = \id A^\iS$;
\item $\Inv(R,S)$ as $S^\iS = \inv{(R^\iS)}$;
\item $\Comp(R,S,T)$ as $T^\iS = R^\iS \comp S^\iS$; and
\item any symbol $\pi'$ in $\{\neq,\notin,\Nrel,\Nset,\Npair\}$ is
   interpreted as $\lnot \pi$ for the corresponding symbol $\pi$ in
   $\{=,\in,\Rel,\set,\pair\}$, where $\lnot$ is logical negation.
\qed
\end{itemize}
\end{definition}

The interpretation structure $\iS$ is used to evaluate each
$\mathcal{RIS}$-formula $\Phi$ into a truth value $\Phi^\iS = \{\true,\false\}$
in the following way: $\mathcal{RIS}$-constraints are evaluated by $\iF{\cdot}$
according to the meaning of the corresponding predicates in set theory as
defined above; $\mathcal{RIS}$-formulas are evaluated by $\iF{\cdot}$ according
to the rules of propositional logic.

In particular, observe that equality between two set terms is interpreted as
the equality in $D_\sSet$; that is, as set equality between hereditarily finite
hybrid sets. Such equality is regulated by the standard \emph{extensionality
axiom}, which has been proved to be equivalent, for hereditarily finite sets,
to the following equational axioms \cite{Dovier00}:
\begin{gather*}
\{x, x \plus A\} = \{x \plus A\} \\
\{x, y \plus A\} = \{y, x \plus A\}.
\end{gather*}

\begin{note}
$\LBR$ can be extended to support other set and relational operators definable
by means of suitable $\LBR$ formulas. Dovier et al. \cite{Dovier00} proved that
symbols in $\PiSB$ are sufficient to define constraints implementing the set
operators $\cap$, $\subseteq$ and $\setminus$. Cristi\'a and Rossi extend that
result in \cite{DBLP:journals/jar/CristiaR20} showing that symbols in $\PiSB
\cup \PiRB$ are sufficient to define constraints implementing all the operators
defined in Note~\ref{note:background}.

As any of these constraints can be replaced by its definition, we can
completely ignore the presence of them in $\LBR$ formulas.
\qed
\end{note}

\begin{note}[Negation]
The negated versions of both set and relational constraints can be introduced as
$\LBR$ formulas \cite{Dovier00,DBLP:journals/jar/CristiaR20}.
For example, $\lnot R = \inv{S}$ is introduced as the %RK predicate $\Ninv(R,S)$ defined as: ($n_i$, existentially quantified):
following ${\cal BR}$-formula:
\begin{equation*}
%RK \Ninv(R,S) \defs
\begin{split}
      ((x&,y) \in R \land (y,x) \notin S)
     {}\lor{} ((x,y) \notin R \land (y,x) \in S) \\
      &{}\lor{} \Nrel(R) \lor \Nrel(S)
\end{split}
\end{equation*}
($x$ and $y$ are implicitly existentially quantified). Thanks to the
availability of negative constraints, (general) logical negation is not
strictly necessary in $\LBR$.
\qed
\end{note}

\vspace{5cm}

\noindent
\textsc{Data Availability Statement} The data underlying this article are available in Dropbox at \url{https://www.dropbox.com/s/
c6z45thxlvr1q1h/setlogITP.zip?dl=0}, and can be accessed with the URL just given.
\end{document}